\documentstyle[12pt,fleqn]{article}
\input{psfig}

\newcommand{\Integer}{\:\mbox{\sf Z} \hspace{-0.82em} \mbox{\sf Z}\,}

\newenvironment{eqar}{
	\setlength{\mathindent}{0 cm}\begin{eqnarray}}{\end{eqnarray}}

\def\nsection#1{\setcounter{equation}{0}\section{#1}}

\def\W#1#2#3#4{W\left(\hspace{-1mm}
         \begin{array}{cc}#4 & #3 \\ #1 & #2 \end{array}
         \hspace{-1mm}\right)}
\def\Wp#1#2#3#4{W'\left(\hspace{-1mm}
         \begin{array}{cc}#4 & #3 \\ #1 & #2 \end{array}
         \hspace{-1mm}\right)}
\def\Wpp#1#2#3#4{W''\left(\hspace{-1mm}
         \begin{array}{cc}#4 & #3 \\ #1 & #2 \end{array}
         \hspace{-1mm}\right)}
\def\Wl#1#2#3#4{W\left(
         #1 \; {{\displaystyle #4 \atop }
          \atop { \atop \displaystyle #2}} \; #3 \right)}
\def\Wlp#1#2#3#4{W'\left(
         #1 \; {{\displaystyle #4 \atop }
          \atop { \atop \displaystyle #2}} \; #3 \right)}
\def\Wlpp#1#2#3#4{W''\left(
         #1 \; {{\displaystyle #4 \atop }
          \atop { \atop \displaystyle #2}} \; #3 \right)}
\def\la{\lambda}

\def\te{\vartheta_1}

\def\tv{\vartheta_4}

\begin{document}

\title{Solvable lattice models labelled by Dynkin diagrams}

\author{S. Ole Warnaar\thanks{e-mail: warnaar@phys.uva.nl}
and Bernard Nienhuis\thanks{e-mail: nienhuis@phys.uva.nl}\\
Instituut voor Theoretische Fysica \\ Universiteit van Amsterdam \\
Valckenierstraat~65 \\ 1018~XE Amsterdam \\ The Netherlands }

\date{ITFA 93-01}

\maketitle

\begin{abstract}
An equivalence between generalised
restricted solid-on-solid (RSOS) mo\-dels,
associated with sets of graphs,
and multi-colour loop models is established.
As an application we consider solvable loop models
and in this way obtain new solvable families of critical RSOS models.
These families can all be classified
by the Dynkin diagrams of the simply-laced Lie algebras.
For one of the RSOS models, labelled by the Lie algebra pair
(A$_L$,A$_L$) and related to the C$_2^{(1)}$ vertex model, we give
an off-critical extension, which breaks the Z$_2$ symmetry of the
Dynkin diagrams.
\end{abstract}
\thispagestyle{empty}
\newpage

\setcounter{page}{1}
\nsection{Introduction}
In recent years many solutions to the star-triangle or Yang-Baxter
equation (YBE) \cite{Baxter} have been found.
Among these solutions, the A-D-E lattice models, found by Pasquier
\cite{Pasquier}, have drawn particular attention.
Pasquier showed, in fact, that to any arbitrary graph one can
associate a solvable restricted solid-on-solid (RSOS) model.
Requiring criticality led to the restriction to graphs
which are Dynkin diagrams
of the simply-laced A-D-E Lie algebras.
An important feature of the A-D-E models is that they can all
be mapped onto the same polygon or loop model,
which in turn is equivalent to the 6-vertex model.

Recently, by extending Pasquier's and similar methods of Owczareck and
Baxter \cite{Owczarek}, Warnaar {\em et al.} \cite{WNS} found a new
family of models associated with graphs. Via a different approach
these same models were also found by Roche \cite{Roche}, who
suggested the name dilute A-D-E models.
Again, the whole family of dilute A-D-E models
can be mapped onto a single loop
model, the O$(n)$ model \cite{Nienhuis}, which is related to the
19-vertex vertex model of Izergin and Korepin \cite{Izergin}.

In this paper we further exploit the relation between RSOS models
related to graphs and loop models.
We define a general multi-colour loop (MCL) model and show its
equivalence with a RSOS model
defined by arbitrary sets of graphs.
Then we consider several examples for which these models are actually
solvable, and find, besides the known A-D-E and dilute A-D-E models,
new families of critical RSOS models labelled by Dynkin diagrams.
As a further generalisation we also consider models of mixed
loop-vertex type.
Finally, for one of the examples, related to the C$_2^{(2)}$ Lie algebra,
we present an off-critical extension.
This extension has the property that it breaks the Z$_2$ symmetry of
the underlying Dynkin diagrams.
In the Appendix we describe the YBE for loop models
\cite{NWB} and show how it relates to the YBE for the RSOS model.

\begin{figure}[hbt]
\centerline{\psfig{file=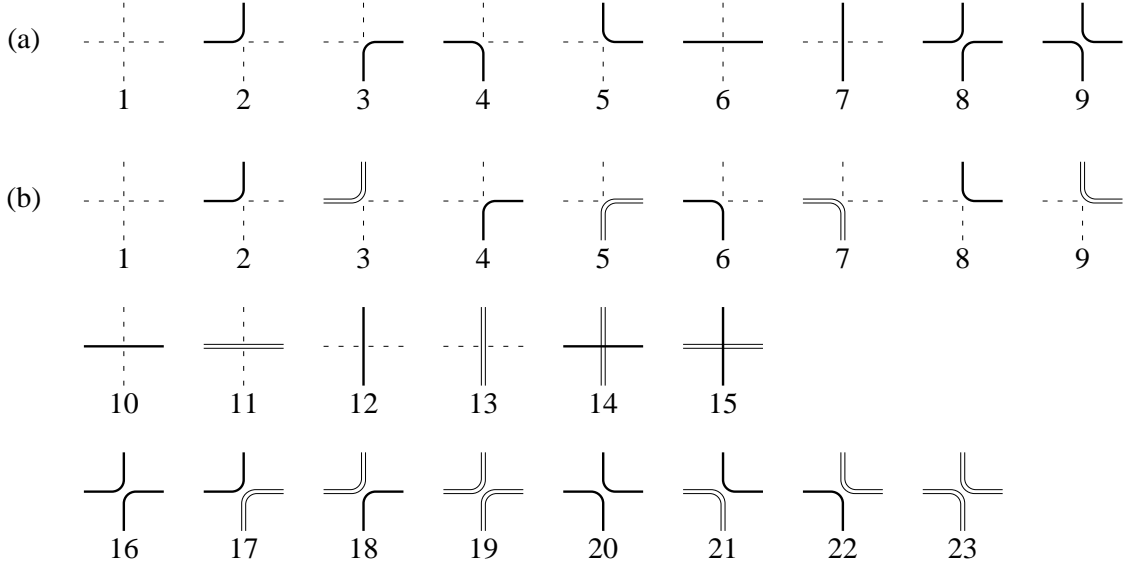,width=14cm}}
\caption{(a) The 9 vertices of the $C=1$ MCL model and (b)
the 23 vertices for the $C=2$ case.}
\label{23vertices}
\end{figure}

\nsection{Multi-colour loop model}
We consider a square lattice $\cal L$.
Each edge of $\cal L$ can either
be occupied by a line segment, that has one of
$C$ possible colours, or be empty.
Line segments of equal colour on adjoining edges must form closed
polygons or loops.
A configuration $G$ is defined as a collection of coloured
loops on $\cal L$, with the restriction that polygons of the same
colour do not intersect. An example of a configuration is given
in figure~\ref{loopsoup}.
The total number of allowed vertices $V$ is given by
$V=3C^2+5C+1$.
For $C=1$ and $C=2$ all possible vertices are shown
in figure~\ref{23vertices}(a) and (b) respectively.
A loop of colour $i$ has fugacity $n_i$ and
the Boltzmann weight of vertex $k$ is given by $\rho_k$.
The partition function of the MCL model is defined as
\begin{equation}
Z = \sum_{G} \rho_1^{m_1} \ldots \rho_V^{m_V}
n_1^{p_1} \ldots n_C^{p_C} ,
\label{loopmodel}
\end{equation}
where $p_i$ is the total number of loops of colour $i$
and $m_k$ the number of vertices of type $k$.
For $C=1$ the MCL model coincides with the loop model defined in \cite{WNS}.

If the weights $\rho_1,\ldots,\rho_{6C+1}$ are all zero,
only configurations that densely cover the entire lattice
give a non-zero contribution to the partition function.
Such loop models we call {\em dense}, opposed to so-called
{\em dilute} loop models which allow for edges of $\cal L$ to
be unoccupied.


\nsection{RSOS model}
In this section we define a restricted solid-on-solid model
and show that its partition function equals that of the multi-colour
loop model.

\subsection{Definition of the model}
Consider an arbitrary connected graph ${\cal G}_i$.
Such a graph consists of a set of $L_i$ nodes,
labelled by an integer {\em height} $a_i\in\{1,\ldots,L_i\}$
and a number of bonds between the nodes.
We do not allow more than one bond between two nodes.
Two nodes are called adjacent $(\sim)$ on ${\cal G}_i$ if
they are connected via a single a bond.
A graph is called simple if it has no
nodes connected to themselves.
Examples of simple and non-simple connected graphs are shown in
figure~\ref{ADEdiagrams} and \ref{tadpoles} respectively.

We can represent the graph ${\cal G}_i$
by an adjacency matrix $A^i$ as follows
\begin{equation}
A_{a_i,b_i}^i=\left\{
\begin{array}{ll}
1 & \qquad a_i \sim b_i \\
0 & \qquad \mbox{otherwise}.
\end{array} \right.
\end{equation}
We denote the largest eigenvalue of $A^i$ by $\Lambda_i$
and the corresponding eigenvector by $S^i$.

We now take $C$ such arbitrary graphs, labelled
${\cal G}_1,\ldots,{\cal G}_C$. Let $a$ be the $C$-di\-men\-si\-o\-nal
vector $a=(a_1,\ldots,a_C)$ and define
\begin{eqnarray}
S_a &=& \prod_i S_{a_i}^i       \nonumber \\
A_{a,b}^i &=& A_{a_i,b_i}^i\prod_{j \neq i} \delta_{a_j,b_j},
\label{factorise}
\end{eqnarray}
where we use the convention that sums and products over $i$ and $j$
always range from 1 to $C$.
With the above definitions the Boltzmann weight of an
elementary face of the RSOS model is defined as
\begin{eqnarray}
\W{a}{b}{c}{d} &=&
\rho_1 \delta_{a,b,c,d}
+ \delta_{a,b,c} \sum_i \rho_{i+1} A_{a,d}^i \nonumber \\
& & \nonumber \\
&+& \delta_{a,c,d} \sum_i \rho_{i+C+1} A_{a,b}^i
+ \left(\frac{S_a}{S_b}\right)^{1/2} \delta_{b,c,d}
\sum_i \rho_{i+2C+1} A_{a,b}^i \nonumber \\
& & \nonumber \\
&+& \left(\frac{S_c}{S_a}\right)^{1/2}
\delta_{a,b,d} \sum_i \rho_{i+3C+1} A_{a,c}^i
+ \delta_{a,b}\delta_{c,d} \sum_i \rho_{i+4C+1}
A_{a,d}^i \label{RSOS} \\
& & \nonumber \\
&+& \delta_{a,d}\delta_{b,c} \sum_i
\rho_{i+5C+1} A_{a,b}^i
+ \sum_{i \neq j} \rho_{k(i,j)} A_{a,d}^i A_{a,b}^j \nonumber \\
& & \nonumber \\
&+& \delta_{a,c} \sum_{i,j} \rho_{l(i,j)}
A_{a,b}^i A_{a,d}^j
+ \left(\frac{S_aS_c}{S_bS_d}\right)^{1/2}
\delta_{b,d} \sum_{i,j} \rho_{m(i,j)} A_{a,b}^i A_{b,c}^j, \nonumber
\end{eqnarray}
where the $i$-th component of the {\em height} vectors $a,b,c$ and $d$
can take any of the
$L_i$ heights on ${\cal G}_i$.
The generalised Kronecker $\delta$ and the functions $k,l$ and $m$
used above are given by
\begin{eqnarray}
\delta_{p,q,\ldots,s}&=&\prod_{i}\delta_{p_i,q_i}
\ldots\delta_{p_i,s_i} \nonumber \\
k(i,j)&=&(C-1)i+j-\theta(j-i)+5C+2 \nonumber \\
l(i,j)&=&Ci+j+C^2+4C+1 \\
m(i,j)&=&Ci+j+2C^2+4C+1, \nonumber
\end{eqnarray}
with $\theta$ the step function:
\begin{equation}
\theta(x)=\left\{
\begin{array}{ll}
0 & x<0 \\ 1 & x \geq 1.
\end{array} \right.
\end{equation}
We note that the total number of terms in equation (\ref{RSOS}) is $V$.

In analogy with the MCL model, if $\rho_1,\ldots,\rho_{6C+1}$ are all
zero, we name the RSOS model dense.
For such models, if all adjacency graphs are simple,
neighbouring sites of the lattice must have
different height.
So-called dilute RSOS models allow for neighbouring sites of $\cal L$
to have equal height.

As will become clear in the following,
we require for dilute RSOS models that all $C$ adjacency
graphs are simple.

\subsection{MCL-RSOS equivalence}\label{RSOSMCL}
We now show that the partition function of the RSOS model, given by
\begin{equation}
Z=\sum_{\mbox{\scriptsize heights}} \prod_{\mbox{\scriptsize faces}}
\W{a}{b}{c}{d},
\label{ZRSOS}
\end{equation}
where the product is over all faces of the square lattice $\cal L$,
can be mapped onto that of the MCL model.
The method is a straightforward generalisation of the
work of Pasquier \cite{Pasquier}, Owczarek and Baxter
\cite{Owczarek} and Warnaar {\em et al.} \cite{WNS}.

As a first step we substitute the expression for the
Boltzmann weight (\ref{RSOS}) and expand the above partition function
into a sum over $V^N$ terms, where $N$ is the number of faces of the
lattice. For each face of $\cal L$, a given term in the expansion has
one of the $V$ terms of equation (\ref{RSOS}).
These $V$ possible terms can be represented diagrammatically
as shown in figure~\ref{23vertices}(a) and (b) for the $C=1$ and
$C=2$ case respectively.
A line of colour $i$,
separating two neighbouring sites with heights
$a$ and $b$ respectively, implies
\begin{eqnarray}
\lefteqn{a_i  \sim  b_i}               \nonumber \\
\lefteqn{a_j  =  b_j \qquad j \neq i.}
\label{restriction}
\end{eqnarray}
We now have to distinguish between dense and dilute RSOS models.

\begin{figure}[hbt]
\centerline{\psfig{file=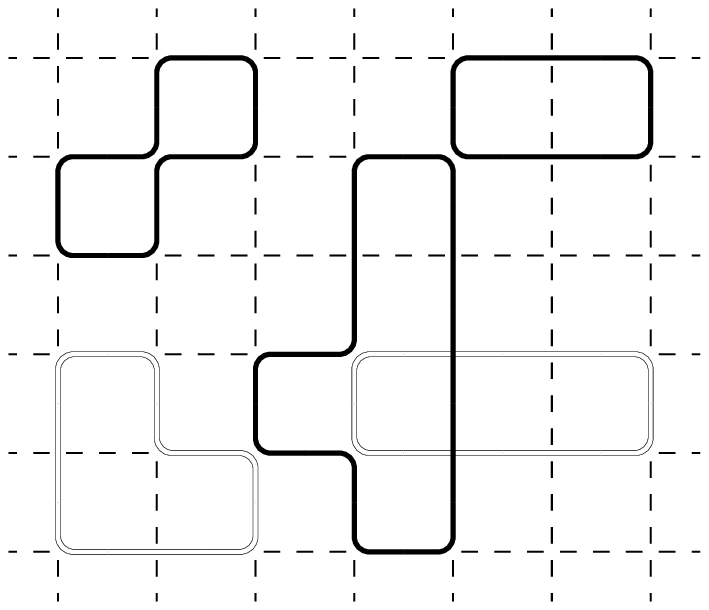}}
\caption{A polygon configuration. Only loops of different colour may
intersect}
\label{loopsoup}
\end{figure}

For dilute RSOS models, since we do not allow for non-simple graphs,
all diagonal elements of the $C$ adjacency matrices are zero.
Consequently, $a_i \sim b_i$ means that $a_i \neq b_i$, and hence
that a line separating two neighbouring sites can be
viewed as a domain wall separating two neighbouring sites with
different heights. As a results of this and the
$\delta$-functions in (\ref{RSOS}),
only configurations in which lines of the same colour join together,
to form domain walls separating
regions of the lattice with different height, give a non-zero
contribution to the partition function.
A typical configuration is shown in figure~\ref{loopsoup}.
If we had allowed for non-simple graphs, configurations where
domain walls would simply end somewhere on the lattice would not give
a vanishing contribution.

For dense RSOS models we do not have this complication.
All edges of the lattice are occupied by polygon segments
and domain walls therefore cannot end.
Besides simple graphs, we can now allow for non-simple adjacency graphs
as well.
If adjacency graph ${\cal G}_i$ is non-simple, a
line of colour $i$ separating two neighbouring sites
does not necessarily separate two sites with different height.
For simplicity we will still refer to such a line as a (local)
domain wall.
Due to the $\delta$-functions we again have that
only configurations in which lines of the same colour join together,
to form global domain walls, give a non-zero
contribution to the partition function.

The partition function is now given as the sum over all configurations
$G$ of domain walls and a sum over heights consistent with $G$
\begin{equation}
Z = \sum_{G} \rho_1^{m_1} \ldots \rho_V^{m_V}
\sum_{\mbox{\scriptsize heights}}
\prod_i \prod_{a_i,b_i=1}^{L_i}
\left(\frac{S_{b_i}^i}{S_{a_i}^i}\right)^{m_{b_i a_i}},
\label{Zexpand}
\end{equation}
where we have used the factorisation property (\ref{factorise})
of $S_a$ and
the meaning of a domain wall of colour $i$, as formulated in
(\ref{restriction}).
The integer $m_{b_i a_i}$ denotes the total power of
$S_{b_i}/S_{a_i}$ arising from the vertices of type
$\rho_{2C+2},\ldots,\rho_{4C+1}$ and $\rho_{2C^2+5C+2},\ldots,\rho_V$,
where we count the powers of
$S_{b_i}/S_{a_i}$ and $S_{a_i}/S_{b_i}$ separately.

To avoid technical difficulties, we assume that all boundary sites of
$\cal L$ carry the same height vector. (For the treatment of other
boundary conditions see \cite{Pasquier,NWB}.)
All domain walls then form closed polygons or loops.
Polygons may of course surround other polygons of any colour,
but can only
be intersected by polygons of a different colour.
If a polygon of colour $i$ is intersected by other polygons,
the height vectors immediately inside and outside
this polygon are not unique.
However, the $i$-th component of these vectors does have a unique
value. We call these the inner and outer height of the polygon
respectively.
We now make the following decomposition:
\begin{eqnarray}
\lefteqn{\hspace{-2.05cm} \psfig{file=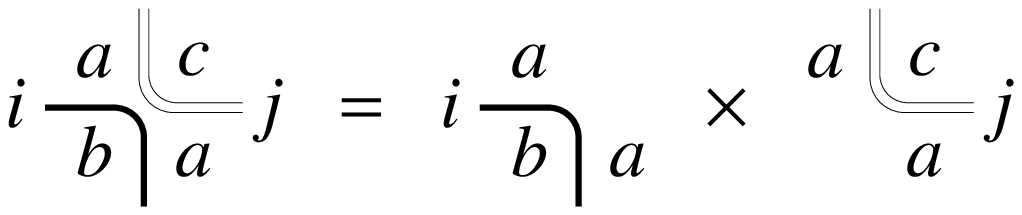,width=6cm}} \nonumber \\
& & \\
\left( \frac{S_{b_i}^i S_{c_j}^j}{S_{a_i}^i S_{a_j}^j}\right)^{1/2}
& &
\left( \frac{S_{b_i}^i}{S_{a_i}^i}\right)^{1/2}
\quad \left( \frac{S_{c_j}^j}{S_{a_j}^j}\right)^{1/2},  \nonumber
\end{eqnarray}
where the labels $i$ and $j$ in the diagrams denote colours.
As a result,
the total contribution to $m_{a_i b_i}-m_{b_i a_i}$
of a polygon of colour $i$, with inner height $a_i$ and outer height
$b_i$ is always 1.

We can now perform the summation over the height vectors in
(\ref{Zexpand}) for each component independently.
When summing over the $i$ height components, we start with polygons of
colour $i$ which do not surround other polygons of the same colour.
If such a polygon has inner height $a_i$ and outer height $b_i$,
we get
\begin{equation}
\sum_{a_i \sim b_i} \frac{S_{a_i}^i}{S_{b_i}^i} = \sum_{a_i=1}^{L_i}
A_{b_i,a_i}^i
\frac{S_{a_i}^i}{S_{b_i}^i}=\Lambda_i.
\label{Lambda}
\end{equation}
The result is that these polygons contribute a factor
$\Lambda_i$ and that their dependence on the outer height $b_i$
disappears. Therefore, the summation over the $i$-th height component
of the regions immediately outside these polygons can now be performed
in the same way.
Repeating this process from inside out and we obtain,
after completely summing out the $i$-th height component,
\begin{equation}
\Lambda_i^{p_i},
\end{equation}
where $p_i$ is the number of polygons  of colour $i$.

If we perform the summation for all $C$ height components, and
make the identification $\Lambda_i=n_i$ we find that
the partition function of the
RSOS model is that of the MCL model.

The equivalence between the partition functions of the MCL and
RSOS model
holds irrespective of the solvability of the models.
Clearly, as a consequence of the equivalence,
if either one of the models is solvable,
in the sense that we can compute its partition function,
the other model is solvable as well.
In the Appendix we show that if the MCL model satisfies the
YBE, then, as an immediate consequence, the YBE for the RSOS model
holds as well.



\nsection{Solvable Examples}
We now consider several special cases
for which the MCL model and hence the RSOS model is solvable.
By a solvable MCL model we mean that it satisfies the
YBE for loop models \cite{NWB}, which is
described in some detail in Appendix~\ref{YBloop}.

Two of these cases are already known in the literature and are
only presented for completeness.
All examples are either 1-, or 2-colour loop models.
So far we have not been able to find any non-trivial solvable
MCL model with more than two colours.

In Appendix~A we give an alternative equivalence between the MCL model
and the RSOS model on the level of the YBE.

\paragraph*{The Temperley-Lieb (TL) loop model}
This dense loop model, which first occurred in the mapping of the
$q$-state self-dual Potts model onto the 6-vertex model \cite{BKW},
is given by equation
(\ref{loopmodel}) with $C=1$.
The weights $\rho_1,\ldots,\rho_9$ of the vertices, shown in
figure~\ref{23vertices}(a), and the fugacity $n_1=\sqrt{q}$ are given by
\begin{equation}
\rho_1=\ldots=\rho_7=0 \qquad \rho_8=\frac{\sin(\la-u)}{\sin \la}
\qquad \rho_9=\frac{\sin u}{\sin \la} \qquad n_1=2\cos \la.
\end{equation}
It is this model for which the equivalence between the RSOS and loop
model was first established \cite{Pasquier,Owczarek}.

\paragraph*{The \boldmath O$(n)$ model}\cite{Nienhuis}
This is a dilute loop model related to the Izergin-Korepin or
A$_2^{(2)}$ vertex model \cite{Izergin}.
It is the most general 1-colour loop model of the form (\ref{loopmodel}).
Again, the dilute RSOS models based on this model have been
constructed before in
\cite{Roche,WNS}.
The Boltzmann weights and fugacity of the O$(n)$ model read
\begin{eqnarray}
\lefteqn{
\rho_1 =
\frac{\sin 2\la \sin 3\la + \sin u \sin(3\la-u)}
{\sin 2\la \sin 3\la}  \qquad
\rho_2 = \rho_3 =
\frac{\sin(3\la-u)}{\sin 3\la} } \nonumber \\
& & \nonumber \\
\lefteqn{
\rho_4 = \rho_5 = \epsilon_1 \: \frac{\sin u}{\sin 3\la} \qquad
\rho_6 = \rho_7 = \epsilon_2 \: \frac{\sin u \sin(3\la-u)}
{\sin 2\la \sin 3\la} } \label{rho} \\
& & \nonumber \\
\lefteqn{
\rho_8 = \frac{\sin(2\la-u) \sin(3\la-u)}
{\sin 2\la \sin 3\la} \qquad
\rho_9 = -\frac{\sin u \sin(\la-u)}
{\sin 2\la \sin 3\la} \qquad n_1 = -2 \cos 4\la, }  \nonumber
\end{eqnarray}
where, here and in the following, $\epsilon_1^2=\epsilon_2^2=1$.

\paragraph*{The C$_2^{(1)}$ loop model}
Whereas the O$(n)$ model is the natural dilute generalisation
of the TL model model,
so the C$_2^{(1)}$ model can be seen as the simplest
non-trivial generalisation of the TL model to a  model with more
than one colour.
It is a dense 2-colour loop model with vertices shown in
figure~\ref{23vertices}(b) and weights
\begin{eqnarray}
\lefteqn{
\rho_1 =\ldots =\rho_{13} = 0
\qquad \rho_{14} = \rho_{15} =
\epsilon_1 \:
\frac{\sin u \sin (3\la-u)}{\sin \la \sin 3\la}} \nonumber \\
& & \nonumber \\
\lefteqn{
\rho_{16} = \rho_{19} =
\frac{\sin(\la-u) \sin(3\la-u)}{\sin \la \sin 3\la} \qquad
\rho_{17} = \rho_{18} = \frac{\sin(3\la-u)}{\sin 3 \la}} \nonumber \\
& & \\
\lefteqn{
\rho_{20} = \rho_{23} =
-\frac{\sin u \sin(2\la-u)}{\sin \la \sin 3\la} \qquad
\rho_{21} = \rho_{22} = \epsilon_2 \:
\frac{\sin u }{\sin 3\la}} \nonumber \\
& & \nonumber \\
\lefteqn{n_1 = n_2 = -2 \cos 2\la.} \nonumber
\end{eqnarray}
By making an arrow covering of the polygons, as described in
section~\ref{mix}, this model maps onto the C$_2^{(1)}$ vertex model of
\cite{Jimbo}.

\paragraph*{The A$_2^{(1)}$ loop model}
This dilute model, related to the A$_2^{(1)}$ vertex model
found in \cite{Perk,Jimbo}, is
given by
(\ref{loopmodel}) with $C=1$ and
\begin{eqnarray}
\lefteqn{
\rho_1 =  \rho_8 = \frac{\sin(\la-u)}{\sin \la}
\qquad \rho_2 = \rho_3 = 1
\qquad \rho_4 = \rho_5 = 0} \nonumber \\
& & \\
\lefteqn{
\rho_6 = \rho_7 = \epsilon_1 \: \frac{\sin u}{\sin \la}
\qquad \rho_9 = \frac{\sin u}{\sin \la}
\qquad n_1=2\cos \la.} \nonumber
\end{eqnarray}

\paragraph*{The A$_3^{(1)}$ loop model}
This dense 2-colour loop model is related to
the A$_3^{(1)}$ vertex model of \cite{Perk,Jimbo}.
The weights are given by
\begin{eqnarray}
\lefteqn{
\rho_1 = \ldots = \rho_{13} = \rho_{21} = \rho_{22} = 0
\qquad \rho_{14} = \rho_{15} = \epsilon_1 \:
\frac{\sin u}{\sin \la}} \nonumber \\
& & \nonumber \\
\lefteqn{
\rho_{16} = \rho_{19} = \frac{\sin (\la-u)}{\sin \la} \qquad
\rho_{17} = \rho_{18} = 1 \qquad
\rho_{20} = \rho_{23} = \frac{\sin u}{\sin \la}} \\
& & \nonumber \\
\lefteqn{n_1=n_2=2\cos \la.} \nonumber
\end{eqnarray}
We remark that, though the A$_1^{(1)}$-A$_3^{(1)}$ vertex
models all relate to loop models, this does not seem to be true for
the A$_n^{(1)}$ family \cite{Perk,Jimbo} in general.


\nsection{Mixed models}\label{mix}
Provided that polygon segments
of the same colour do not intersect, each loop
model can be mapped onto a vertex model \cite{BKW}.
We cover loops of colour $i$ by arrows of that same colour, such
that the loops become oriented. Following a loop in the direction of
the arrows, we assign a phase factor $s_i$ to a turn to the left and
a factor $s_i^{-1}$ to a turn to the right, where $s_i$ is defined
by $n_i=s_i^4+s_i^{-4}$.
Summing over all possible arrow coverings of a configuration,
each polygon of colour $i$ acquires a
total factor $n_i$. Interchanging the summation
over all loop configurations
and over all arrow coverings,
the sum over the loops can readily be performed.
The resulting partition function is that of a vertex model, where the
arrows arround a vertex obey the ice-rule for each colour
independently.

In general, the inverse of the above mapping
is not possible.
Only very few
solvable vertex models, that satisfy an ice-rule, admit a loop
interpretation.
Nevertheless, many vertex models allow for a partial mapping
onto a loop model. That is, some vertex degrees of freedom can be
converted into loop degrees of freedom, but not all.
Via the MCL-RSOS correspondence, these mixed {\em loop-vertex}
models can be mapped onto {\em RSOS-vertex} or equivalently
{\em RSOS-SOS} models.

\begin{figure}[hbt]
\centerline{\psfig{file=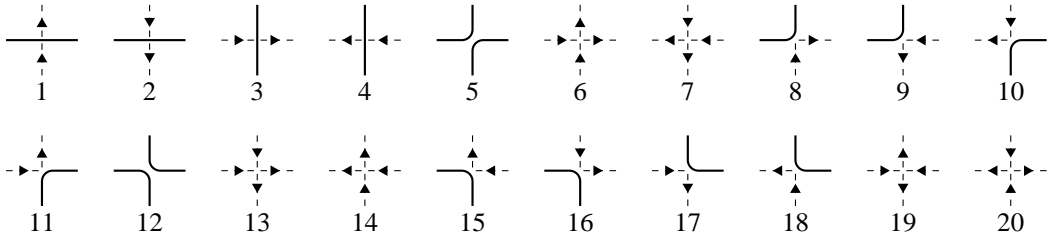,width=14cm}}
\caption{The 20 vertices of the A$_3^{(2)}$ loop-vertex model.}
\label{20vertices}
\end{figure}

We shall not try to give a complete description off all models that
allow such a procedure, but consider as an example the
A$_3^{(2)}$ vertex model.
For the definition of this 36-vertex model we refer to \cite{Jimbo}.
The equivalent loop-vertex model has 20 vertices, shown in
figure~\ref{20vertices}, with weights and fugacity
\begin{eqnarray}
\lefteqn{
\rho_1 = \ldots = \rho_4 =
\epsilon_1 \: \frac{\sin u \cos(2\la-u)}{\sin \la \cos 2\la}
\qquad
\rho_5 = \rho_6 = \rho_7 =
\frac{\sin(\la-u)\cos(2\la-u)}{\sin \la \cos 2\la}
} \nonumber \\
& & \nonumber \\
\lefteqn{
\rho_8 = \ldots = \rho_{11} =
\frac{\cos(2\la-u)}{\cos 2\la}
\qquad
\rho_{12} = \rho_{13} = \rho_{14} =
-\frac{\sin u \cos(\la-u)}{\sin \la \cos 2\la} } \\
& &  \nonumber \\
\lefteqn{
\rho_{15} = \ldots = \rho_{18} =
\epsilon_2 \: \frac{\sin u}{\cos 2\la}
\qquad
\rho_{19} = \rho_{20} = 1
\qquad
n_1= -2 \cos 2\la.
} \nonumber
\end{eqnarray}

Via the MCL-RSOS equivalence, this model maps onto a RSOS-vertex
model, where an elementary face of the lattice is denoted as
\begin{equation}
W\left(\hspace{-1mm}\begin{array}{ccc}
d & \gamma & c \\
\delta & & \beta \\
a & \alpha & b \\
 \end{array} \hspace{-1mm}\right)=
W\left(\hspace{-1mm}\begin{array}{ccc}
d_1 & \gamma & c_1 \\
\delta & & \beta \\
a_1 & \alpha & b_1 \\
 \end{array} \hspace{-1mm}\right)=
\ldots.
\end{equation}
The latin indices label 1-dimensional
height variables and the greek indices
the arrows.
We note that two neighbouring sites either have the same height
separated by an arrow or have different heights.

To cast this into a somewhat nicer form, we make use of the
SOS-vertex equivalence for ice-type models \cite{Beijeren} to
write this as a RSOS-SOS model.
For that purpose we assign 2-dimensional height vectors $a$ to each
site of the lattice.
The first component of such a vector is one of the heights $a_1\in
\{1,\ldots,L_1\}$. The second component is a height variable $a_2\in
\Integer$.
Two neighbouring sites of the RSOS-vertex model, with heights $a_1$
and $b_1$ separated by an arrow,
correspond to height vectors $a$ and $b$ of the RSOS-SOS model, with
\begin{eqnarray}
\lefteqn{a_1=b_1} \nonumber \\
\lefteqn{a_2-b_2=\pm 1,}
\end{eqnarray}
and the
convention that the height to the left of the arrow is
highest.
For two sites of the RSOS-vertex model that are not separated by an
arrow, we get heights $a$ and $b$, with
\begin{eqnarray}
\lefteqn{a_1 \sim b_1} \nonumber \\
\lefteqn{a_2=b_2.}
\end{eqnarray}

If we define $A_{a,b}^i$ $(i=1,2)$
and $S_a$ as in equation (\ref{factorise}), where
\begin{equation}
\begin{array}{l}
A_{a_2,b_2}^2 = \delta_{a_2,b_2-1} + \delta_{a_2,b_2+1} \\
 \\
S_{a_2}^2 = 1
\end{array}\qquad a_2, b_2\in\Integer,
\end{equation}
we finally get for the A$_3^{(2)}$ SOS model
\begin{eqnarray}
\W{a}{b}{c}{d} &=&
\rho_1\sum_{i\neq j}
A_{a,b}^i A_{a,d}^j
+ \delta_{a,c}\left(
\rho_5 \sum_{i=j} \quad
+\rho_8\sum_{i\neq j} \quad \right)
A_{a,b}^i A_{a,d}^j \nonumber \\
& & \nonumber \\
&+&\delta_{b,d} \left(\frac{S_a S_c}{S_b S_d}\right)^{1/2} \left(
\rho_{12} \sum_{i=j} \quad
+\rho_{15} \sum_{i\neq j} \quad \right)
A_{a,b}^i A_{b,c}^j  \\
& & \nonumber \\
&+&\left(\rho_{19}-\rho_5-\rho_{12}
\right)
\delta_{a,c}\delta_{b,d} A_{a,b}^2 A_{a,d}^2. \nonumber
\end{eqnarray}

Similarly, we can construct mixed SOS models starting from
other vertex models. For the solutions of the YBE
found in \cite{Jimbo} we get
models with height variables that have $r$ restricted and $u$
unrestricted components, with $r$ and $u$ listed below
\begin{equation}
\begin{array}{lll}
\mbox{A}_{2n-1}^{(1)}: & r=2&u=n-2 \quad (n \geq 2)\\
\mbox{A}_{2n}^{(1)}: & r=1&u=n-1 \\
\mbox{C}_n^{(1)}: & r=2 & u=n-2 \quad  (n \geq 2) \\
\mbox{A}_{2n-1}^{(2)}: & r=1 & u=n-1  \\
\mbox{A}_{2n}^{(2)}: & r=1 & u=n-1.  \\
\end{array}
\end{equation}

\begin{figure}[hbt]
\centerline{\psfig{file=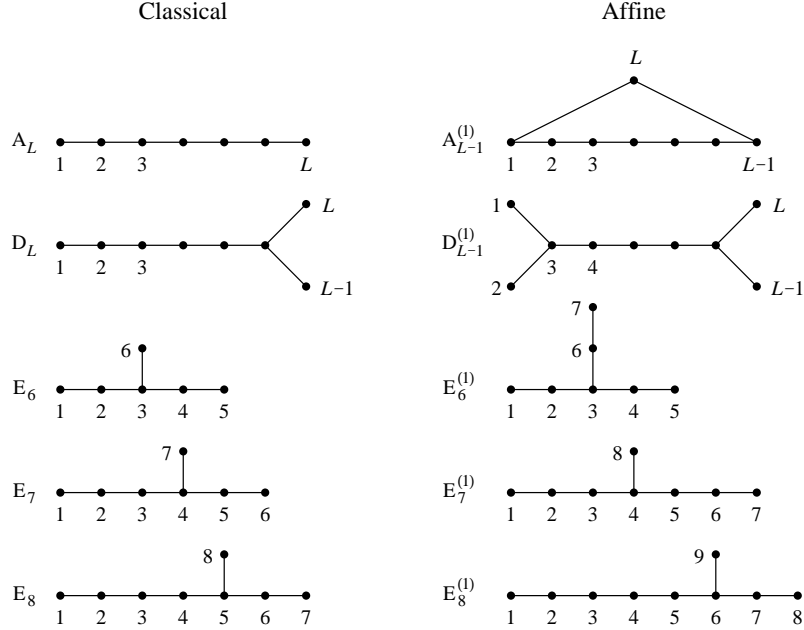,height=8.5cm}}
\caption{Dynkin diagrams of the simply-laced Lie algebras.}
\label{ADEdiagrams}
\end{figure}
\begin{figure}[hbt]
\centerline{\psfig{file=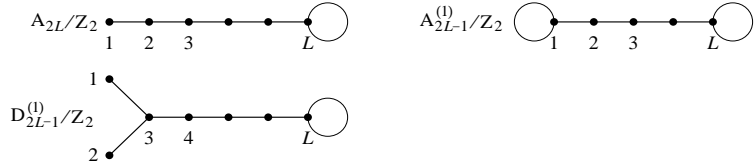,width=10cm}}
\caption{Non-simple connected graphs with
largest eigenvalue $\leq 2$.}
\label{tadpoles}
\end{figure}

\nsection{A-D-E classification}
All loop models presented in the previous sections
are critical when $n_i\leq 2$.
For $n_i>2$ the trigonometric functions have to be replaced by
hyperbolic functions and the
models become non-critical and in some cases even complex.
It is therefore natural to restrict the graphs ${\cal G}_i$ of the
RSOS models, to those
that have adjacency matrices $A^i$ with largest eigenvalue $\Lambda_i$
less or equal than two.
In fact, all simple connected graphs
with $\Lambda_i\leq 2$ have been classified \cite{Cvetkovic}
and are given by
the Dynkin diagrams of the classical and affine simply-laced
Lie algebras, shown in figure~\ref{ADEdiagrams}.
For the classical algebras,
the largest eigenvalue of the adjacency matrix of its Dynkin diagram
is given by $2\cos\frac{\pi}{h}$, where $h$ is the Coxeter number of the
algebra. For the affine algebras the largest eigenvalue is 2.
The respective values of $h$, and the
largest eigenvectors are given in table~\ref{thetable} for each algebra.
\begin{table}[hbt]
\centering
$$
\begin{array}{|c|c|c|}
\hline
\mbox{algebra} & h & \mbox{Perron-Frobenius vector} \\
\hline
\mbox{A}_L & L+1 &
(\sin\frac{\pi}{h},\sin\frac{2\pi}{h},\ldots,\sin\frac{L\pi}{h}) \\
\mbox{D}_L & 2L-2 &
(2\sin\frac{\pi}{h},\ldots,2\sin\frac{(L-2)\pi}{h},1,1) \\
\mbox{E}_6 & 12 &
(\sin\frac{\pi}{h},\ldots,\sin\frac{3\pi}{h},
\frac{\textstyle \sin\frac{5\pi}{h}}{\textstyle 2\cos\frac{\pi}{h}},
2\cos\frac{4\pi}{h} \sin\frac{\pi}{h},
\frac{\textstyle \sin\frac{3\pi}{h}}{\textstyle 2\cos\frac{\pi}{h}}) \\
\mbox{E}_7 & 18 &
(\sin\frac{\pi}{h},\ldots,\sin\frac{4\pi}{h},
\frac{\textstyle \sin\frac{6\pi}{h}}{\textstyle 2\cos\frac{\pi}{h}},
2\cos\frac{5\pi}{h} \sin\frac{\pi}{h},
\frac{\textstyle \sin\frac{4\pi}{h}}{\textstyle 2\cos\frac{\pi}{h}}) \\
\mbox{E}_8 & 30 &
(\sin\frac{\pi}{h},\ldots,\sin\frac{4\pi}{h},
\frac{\textstyle \sin\frac{7\pi}{h}}{\textstyle 2\cos\frac{\pi}{h}},
2\cos\frac{6\pi}{h} \sin\frac{\pi}{h},
\frac{\textstyle \sin\frac{5\pi}{h}}{\textstyle 2\cos\frac{\pi}{h}}) \\
\mbox{A}_{L-1}^{(1)} & L &
(1,1,\ldots,1)  \\
\mbox{D}_{L-1}^{(1)} & 2L-6 &
(1,1,2,\ldots,2,1,1) \\
\mbox{E}_{6}^{(1)} & 6 &
(1,2,3,2,1,2,1) \\
\mbox{E}_{7}^{(1)} & 12 &
(1,2,3,4,3,2,1,2) \\
\mbox{E}_{8}^{(1)} & 30 &
(1,2,3,4,5,6,4,2,3) \\
\hline
\end{array}
$$
\caption{Coxeter number and largest eigenvector of the simply-laced
Lie algebras}
\label{thetable}
\end{table}
In the case of RSOS models, we also alow for non-simple
connected graphs.
However, none of the graphs with largest eigenvalue $\leq 2$,
leads to intrinsically new models,
as they can always be viewed as one of the simple graphs where
a Z$_2$ symmetry is modded out, see
figure~\ref{tadpoles}.

\nsection{Off-critical models}
It is well-known that all critical A-D models
based on the Temperley-Lieb loop model admit
an extension away from criticality while remaining solvable.
The off-critical models based on the classical Lie algebra A$_L$
for example, are the models of Andrews, Baxter and Forrester
\cite{ABF}.

Recently, the off-critical extension of the dilute A$_L$ models
based on the
O$(n)$ model was found \cite{WNS}. As an interesting feature, these
models break a Z$_2$ symmetry of the underlying Dynkin diagrams,
when $L$ is odd.

A natural question therefore is: which of the new A-D-E models
presented in section~4,
admit an extension away from criticality?
So far, we have not studied this problem in any systematic way.
However, for the C$_2^{(1)}$ RSOS model based on the Lie algebra
pair (A$_L$, A$_L$), the extension can easily be found by making the
transformation $a_1\la\rightarrow a_1\la +\frac{1}{2}i\ln p$
to the C$_2^{(1)}$ RSOS model of
Jimbo {\em et al.} \cite{JimboMiwaOkado}.

This solutions involves the $\vartheta$-functions \cite{GR}
\begin{eqnarray}
\te(u) &=& 2 p^{1/4} \sin u \prod_{n=1}^{\infty}
(1-2 p^{2n} \cos 2u+p^{4n})(1-p^{2n}) \nonumber \\
\tv(u) &=& \prod_{n=1}^{\infty}
(1-2p^{2n-1} \cos 2u+p^{4n-2})(1-p^{2n}),
\end{eqnarray}
where we have suppressed the dependence on the nome $p$,
$|p|<1$.
If we also define the unit-vectors $e_1=(1,0)$ and $e_2=(0,1)$, and
use the notation $e_{-\mu}=-e_{\mu}$, $a_{-\mu}=-a_{\mu}$,
$\mu=\pm 1$, $\pm 2$, where $a=(a_1,a_2)$,
the solution reads
\begin{eqar}
\lefteqn{\W{a}{a-e_{\mu}}{a}{a+e_{\mu}}=
\frac{\te(\la-u)\te(3\la-u)}{\te(\la)\te(3\la)}}
\nonumber \\  & & \nonumber \\
\lefteqn{\W{a+e_{\mu}}{a}{a-e_{\mu}}{a}=
-\left(\frac{S(a+e_{\mu})S(a-e_{\mu})}{S^2(a)}\right)^{1/2}
\frac{\te(u)\te(2\la-u)}{\te(\la)\te(3\la)}}
\nonumber \\ & & \nonumber \\
\lefteqn{\W{a}{a-e_{\nu}}{a}{a+e_{\mu}}=
\frac{\te(3\la-u)\tv(a_{\mu}\la-a_{\nu}\la+\la-u)}
{\te(3\la)\tv(a_{\mu}\la-a_{\nu}\la+\la)}}
\nonumber \\ & & \nonumber \\
\lefteqn{\W{a+e_{\mu}}{a}{a-e_{\nu}}{a}=\epsilon_2
\left(\frac{S(a+e_{\mu})S(a-e_{\nu})}{S^2(a)}\right)^{1/2}
\frac{\te(u)\tv(a_{\mu}\la-a_{\nu}\la-2\la+u)}
{\te(3\la)\tv(a_{\mu}\la-a_{\nu}\la+\la)}}
\nonumber \\ & & \nonumber \\
\lefteqn{\W{a+e_{\mu}}{a}{a+e_{\mu}}{a}=
\frac{S(a+e_{\mu})}{S(a)}
\frac{\te(u)\te(2a_{\mu}\la-2\la+u)}{\te(3\la)\te(2a_{\mu}\la+\la)}}
\\ & & \nonumber \\
\lefteqn{\hspace{7cm}
+\frac{\te(3\la-u)\te(2a_{\mu}\la+\la+u)}{\te(3\la)\te(2a_{\mu}\la+\la)}}
\nonumber \\ & & \nonumber \\
\lefteqn{\W{a+e_{\mu}}{a+e_{\mu}+e_{\nu}}{a+e_{\nu}}{a}}
\nonumber \\ & & \nonumber \\
\lefteqn{\hspace{2cm}
=\epsilon_1
\left(\frac{\tv(a_{\mu}\la-a_{\nu}\la-\la)\tv(a_{\mu}\la-a_{\nu}\la+\la)}
           {\tv^2(a_{\mu}\la-a_{\nu}\la)}\right)^{1/2}
\frac{\te(u)\te(3\la-u)}{\te(\la)\te(3\la)}}
\nonumber \\ & & \nonumber \\
\lefteqn{S(a)=(-)^{\displaystyle a_1+a_2} \;
\te(2a_1\la)\te(2a_2\la)\tv(a_1\la-a_2\la)\tv(a_1\la+a_2\la),}
\nonumber
\end{eqar}
where $\nu \neq \pm \mu$ and $\la=\frac{\pi}{2}\frac{L}{L+1}$.
Like the ABF model there are four different physical regimes:
\begin{equation}
\begin{array}{cll}
\left. \begin{array}{c} 0<p<1 \\ -1<p<0 \end{array} \right\}
&\quad 0<u<3\la-\pi  &\qquad \epsilon_1=-\epsilon_2=1 \\
& & \\
\left. \begin{array}{c} 0<p<1 \\ -1<p<0 \end{array} \right\}
&\quad 3\la-2\pi<u<0  &\qquad \epsilon_1=-\epsilon_2=-1.
\end{array}
\end{equation}
Due to periodicity of the weights,
the choice $\la=\frac{\pi}{2}\frac{L+2}{L+1}$, which also yields
positive largest eigenvalues $\Lambda_1=\Lambda_2$, does not
give any new regimes.

We note that, away from criticality, the above models break the Z$_2$
symmetry of the underlying dynkin diagrams, when $L$ is odd.
If we define $\tilde{a}_i=L+1-a_i$, we have
\begin{equation}
W \left(
\begin{array}{cc}
(d_1,d_2) & (c_1,c_2) \\
(a_1,a_2) & (b_1,b_2)
\end{array}
\right) \neq
W \left(
\begin{array}{cc}
(\tilde{d}_1,d_2) & (\tilde{c}_1,c_2) \\
(\tilde{a}_1,a_2) & (\tilde{b}_1,b_2)
\end{array}
\right) \qquad L\mbox{ odd}.
\end{equation}
and similarly for the second component. We do however retain the
symmetry
\begin{equation}
W \left(
\begin{array}{cc}
(d_1,d_2) & (c_1,c_2) \\
(a_1,a_2) & (b_1,b_2)
\end{array}
\right) =
W \left(
\begin{array}{cc}
(\tilde{d}_1,\tilde{d}_2) & (\tilde{c}_1,\tilde{c}_2) \\
(\tilde{a}_1,\tilde{a}_2) & (\tilde{b}_1,\tilde{b}_2)
\end{array}
\right).
\end{equation}


\nsection{Summary and discussion}
We have established a graphical equivalence between restricted
solid-on-solid models and loop models.
In particular, we have applied this equivalence to solvable
loop models and, as a result, found new families of critical RSOS
models. These new models can all be classified in terms of
Dynkin diagrams
of the simply-laced Lie algebras, the so-called A-D-E algebras.
Furthermore we have indicated how to extend the equivalence to
models that are of mixed loop-vertex type.
Finally, an off-critical extension of the C$_2^{(1)}$
RSOS model based on the Dynkin diagram pair (A$_L$,A$_L$) is given. This
extension, which involves elliptic $\vartheta$-functions, breaks the
Z$_2$ symmetry off the underlying Dynkin diagrams.

Obvious generalisations of the ideas presented in this paper are:
\begin{description}
\item[{\rm (i)}]
     The extension to directed adjacency graphs, see {\em e.g.} \cite{DiF}.
\item[{\rm (ii)}]
      The study of loop models that admit multiple occupation of
      edges. That is, each edge of the lattice can be occupied by
      more that one polygon segment, provided that all segments
      have different colour. Clearly the MCL-RSOS equivalence of
      section~\ref{RSOSMCL} still holds.
\item[{\rm (iii)}]
      The extension to higher spin RSOS models,
      where we view the dense and dilute RSOS models
      as spin-$\frac{1}{2}$ and spin-1 models respectively.
\item[{\rm (iv)}]
     The mapping of loop models onto RSOS-vertex or RSOS-SOS models.
     In section~\ref{RSOSMCL}
     we have shown how a loop model can be mapped onto
     a RSOS model by identifying the fugacity $n_i$
     of a loop of colour $i$ with the largest eigenvalue
     $\Lambda_i$ of an adjacency matrix $A^i$.
     In section~\ref{mix}
     we have shown how a loop model can be mapped onto
     a vertex model by setting $n_i = s_i^4 + s_i^{-4}$, where the
     phase factor $s_i$ $(s_i^{-1})$
     is associated with a directed loop
     making a turn to the left (right).
     Combining these two mapping, choosing
     $n_i = \Lambda_i (s_i^4 + s_i^{-4})$, we can map a loop model
     on a RSOS-vertex model or RSOS-SOS model. We note that this
     type of RSOS-SOS model is altogether different from the
     RSOS-SOS models defined in section~\ref{mix}.
\end{description}
We hope to report a study of these generalisations in a
future publications.

An intriguing open problem \cite{JimboMiwaOkado}
is the relation between the critical
A$_n^{(1)}$, B$_n^{(1)}$, C$_n^{(1)}$, D$_n^{(1)}$, A$_{2n}^{(2)}$
and A$_{2n-1}^{(2)}$ RSOS models found in \cite{JimboMiwaOkado}
and \cite{Kuniba}
and their
vertex couterparts given in \cite{Jimbo}.
For some of these models, notably the A$_1^{(1)}$, C$_2^{(1)}$ and
A$_{2}^{(2)}$
models, the MCL-RSOS equivalence, does provide a link between the RSOS and
vertex representations.
It remains unclear however, how to extend the methods of this paper to
establish the RSOS-vertex correspondence in general.


\section*{Acknowledgements}
We thank Paul A. Pearce for kind interest in our work.
This work has been supported by the Stichting voor
Fundamenteel Onderzoek der Materie (FOM).

\appendix
\nsection{Yang-Baxter equation for the MCL and RSOS model}
The equivalence between the MCL and RSOS model holds irrespective of
the solvability of the models.
In this appendix we show however that a sufficient condition for the
YBE equation of the RSOS model to hold, is that the corresponding
MCL model satisfies the YBE.

\subsection{Yang-Baxter equation for loop models}\label{YBloop}
Although loop models are intrinsically non-local, one can nevertheless
formulate a local condition or YBE for two transfer matrices to
commute \cite{NWB}. (For the definition of the transfer matrix for
loop models, see {\em e.g.} \cite{BN}.)
In order to define this equation we need some preliminaries.

\begin{figure}[h]
\centerline{\psfig{file=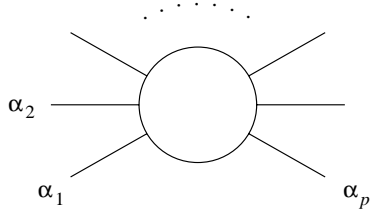,width=5cm}}
\caption{An object with $p$ external edges.}
\label{plegs}
\end{figure}

Consider an object $\cal O$ with $p$ external edges, labelled
$1,\ldots,p$, as shown in figure~\ref{plegs}.
Edges of $\cal O$ can either be empty or occupied by a coloured
polygon segment.
Each edge that is occupied is, via the interior of $\cal O$,
connected to a one other edge that is occupied by a line segement
of equal colour.
The index $\alpha_k$ contains the following information:
(i) whether edge $k$ is occupied by a polygon segment of given colour,
and (ii) if so, to which other edge it is connected.
The information contained in all
$\alpha_1,\ldots,\alpha_p$ is called the
{\em connectivity} of ${\cal O}={\cal O}(\alpha_1,\ldots,\alpha_p)$
and denoted by $C_{\cal O}$.
The object $\cal O$ has a weight $W({\cal O})$.

By the contraction of $\cal A$ and $\cal B$ to the composite object
$\cal D$
\begin{eqnarray}
\lefteqn{\sum_{C_{\cal A},C_{\cal B}}
{\cal A}(\alpha_1,\ldots,\alpha_k,\mu_1,\ldots,\mu_l)
{\cal B}(\beta_1,\ldots,\beta_m,\mu_1,\ldots,\mu_l)} \nonumber \\
& & \hspace{6cm}=
{\cal D}(\alpha_1,\ldots,\alpha_k,\beta_1,\ldots,\beta_m),
\end{eqnarray}
we glue together the edges of $\cal A$ and $\cal B$ that carry
the same index and sum over all connectivities of $\cal A$ and
$\cal B$ consistent with $C_{\cal D}$.
Here it is to be understood that $\mu_i$ in the argument of $\cal A$
and of $\cal B$ still signify two different things. We use the repeated
occurence of the labels only to indicate that the edge of $\cal A$
that carries the index $\mu_i$ is glued to the edge of $\cal B$ that
carries that same index. Furthermore it implies that edges which are
glued together must be occupied by a polygon segment of the same colour.
Finally it means that if the edge of $\cal A$
($\cal B$) carrying the index
$\mu_i$ is occupied and connected to, say, the edge of $\cal A$
($\cal B)$ carrying the index $\alpha_j$ ($\beta_k)$ then
then the edges of $\cal D$ carrying the indices
$\alpha_j$ and $\beta_k$ are connected.

The weight $W({\cal D})$ is defined as
\begin{equation}
W({\cal D}) \sum_{C_{\cal A},C_{\cal B}} W({\cal A}) W({\cal B})
n_1^{p_1} \ldots n_C^{p_C},
\end{equation}
where $p_i$ is the number of polygons of colour $i$
that are closed by glueing together
$\cal A$ and $\cal B$.

\begin{figure}[h]
\centerline{\psfig{file=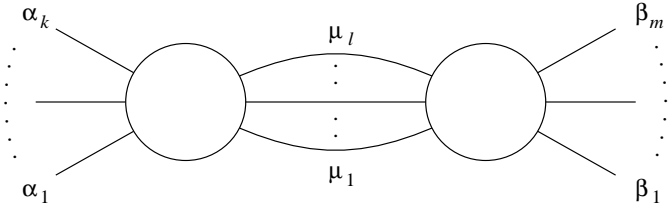,width=9cm}}
\caption{Graphical representation of the
contraction of $\cal A$ and $\cal B$.}
\label{contraction}
\end{figure}

An elementary vertex of the MCL model has four external edges.
Because the Boltzmann weight $W$ of a vertex $\cal V$
is completely determined
by its connectivity, we can write
\begin{equation}
W = W({\cal V}(\alpha,\beta,\gamma,\delta)) \equiv
\Wl{\alpha}{\beta}{\gamma}{\delta}.
\end{equation}
With the above definitions, the YBE equation for the MCL model may be
written as:
\begin{eqnarray}
\lefteqn{\sum_{C_{\cal V},C_{\cal V'},C_{\cal V''}}
\Wl{\alpha}{\beta}{\mu}{\nu}
\Wlp{\zeta}{\nu}{\tau}{\epsilon}
\Wlpp{\tau}{\mu}{\gamma}{\delta}} \nonumber \\
& &  \\
\lefteqn{\qquad \qquad \qquad \qquad =
\sum_{C_{\cal V},C_{\cal V'},C_{\cal V''}}
\Wlpp{\zeta}{\alpha}{\tau}{\mu}
\Wlp{\tau}{\beta}{\gamma}{\nu}
\Wl{\mu}{\nu}{\delta}{\epsilon},} \nonumber
\end{eqnarray}
and must be satisfied for all possible conectivities
$C_{\mbox{\scriptsize YBE}(\alpha,\ldots,\zeta)}$.
In other words, not only do we fix the occupation of the external
edges,
but also to which other external edge an occupied edge
is connected.
If one of the terms in the above equation has an internal loop of
colour $i$ this yields a factor $n_i$.

\subsection{Yang-Baxter equation for the RSOS model}
To show that if the
YBE for the MCL model is satisfied, it holds as well for the RSOS
model,
we begin with the YBE for the RSOS model \cite{Baxter}
\begin{eqnarray}
\lefteqn{
\sum_g \W{a}{b}{g}{f} \Wp{f}{g}{d}{e} \Wpp{g}{b}{c}{d}} \nonumber \\
& & \\
\lefteqn{\qquad \qquad \qquad \qquad
=\sum_g \Wpp{f}{a}{g}{e} \Wp{a}{b}{c}{g}
\W{g}{c}{d}{e} .} \nonumber
\end{eqnarray}
This equation must hold for all values of the external height vectors,
with $a_i,\ldots,g_i \in \{1,\ldots,L_i\}$.
We substitute the definition of the weights $W$, $W'$ and
$W''$, where $W'$ and $W''$ are given by (\ref{RSOS}) with $\rho_k$
replaced by $\rho_k'$ and $\rho_k''$, and expand
both sides of the YBE into $V^3$ terms.
We then use the factorisation (\ref{factorise}) and
perform the trivial summation over the $\delta$-functions.
As a result, most terms in the expansion no longer contain
the variable $g_i$.
Only terms for which the internal site differs in height from all its
three neighbouring sites,
which in that case all have equal height, yield a $g_i$ dependent
factor of the form $S_{g_i}/S_{a_i}$. Here $g_i$ and $a_i$ are the
$i-$th height components of the center site and its neighbouring sites
respectively, and $a_i\sim g_i$, $a_j=g_j$ $j\neq i$.
Performing the sum over $g_i$ yields a factor $\Lambda_i$, see
equation (\ref{Lambda}).
We now group together all terms that have the same dependence
on the vectors $S^i$.
If we demand that the resulting equation holds for any arbitrary set
of graphs $ \{{\cal G}_1,\ldots,{\cal G}_c\}$, a sufficient and
presumably necessary condition is that
each group of terms vanishes independently.
If we draw domain walls (of the appropriate colour)
between regions of different height, terms within the same group
all have the same connectivity. Furthermore, a term with an internal loop
of colour $i$ contributes an extra factor $\Lambda_i$.
As a result we find, upon setting $\Lambda_i=n_i$, that each group
yields precisely one of the equations of the YBE for the MCL model.



\newpage

\listoffigures
\end{document}